\documentclass[twocolumn,floatfix,prb,showpacs]{revtex4}
\usepackage{graphicx}
\usepackage{amsmath}
\usepackage{bm}
\usepackage{hyperref}
\usepackage{soul}
\usepackage{ulem}
\usepackage{lipsum}
\usepackage{color}
\begin{document}
\title{Andreev reflection in monolayer MoS$_2$ }

\author{Leyla Majidi}
\email{leyla.majidi@ipm.ir}
\affiliation{School of Physics, Institute for Research in Fundamental Sciences (IPM), Tehran 19395-5531, Iran}
\author{Habib Rostami}
\affiliation{School of Physics, Institute for Research in Fundamental Sciences (IPM), Tehran 19395-5531, Iran}
\author{Reza Asgari}
\email{asgari@ipm.ir}
\affiliation{School of Physics, Institute for Research in Fundamental Sciences (IPM), Tehran 19395-5531, Iran}

%%%%%%%%%%%%%%%%%%%%%%%%%%%%%%%%%%%%%%%%%%%%%%%%%%%%%%%%%%%%%%%%%%%%%%%%%%%%%

\begin{abstract}
Andreev reflection in a monolayer molybdenum disulfide superconducting-normal (S/N) hybrid junction is investigated. We find, by using a modified-Dirac Hamiltonian and the scattering formalism, that the perfect Andreev reflection happens at normal incidence with $p$-doped S and N regions. The probability of the Andreev reflection and the resulting Andreev conductance, in this system, are demonstrated to be large in comparison with the corresponding gapped graphene structure. We further investigate the effect of a topological term ($\beta)$ in the Hamiltonian and show that it results in an enhancement of the Andreev conductance with $p$-doped S and N regions, while in the corresponding structure with $n$-doped S region it is strongly reducible in comparison. This effect can be explained in terms of the dependence of the Andreev reflection probability on the sign of $\beta$ and the chemical potential in the superconducting region.

\end{abstract}
%%%%%%%%%%%%%%%%%%%%%%%%%%%%%%%%%%%%%%%%%%%%%%%%%%%%%%%%%%%%%%%%%%%%%%%%%%%

\maketitle

%%%%%%%%%%%%%%%%%%%%%%%%%%%%%%%%%%%%%%%%%%%%%%%%%%%%%%%%%%%%%%%%%%%%%%%%%%%
\section{\label{sec:intro}Introduction}
\begin{figure}[t]
\begin{center}
\includegraphics[width=1\linewidth]{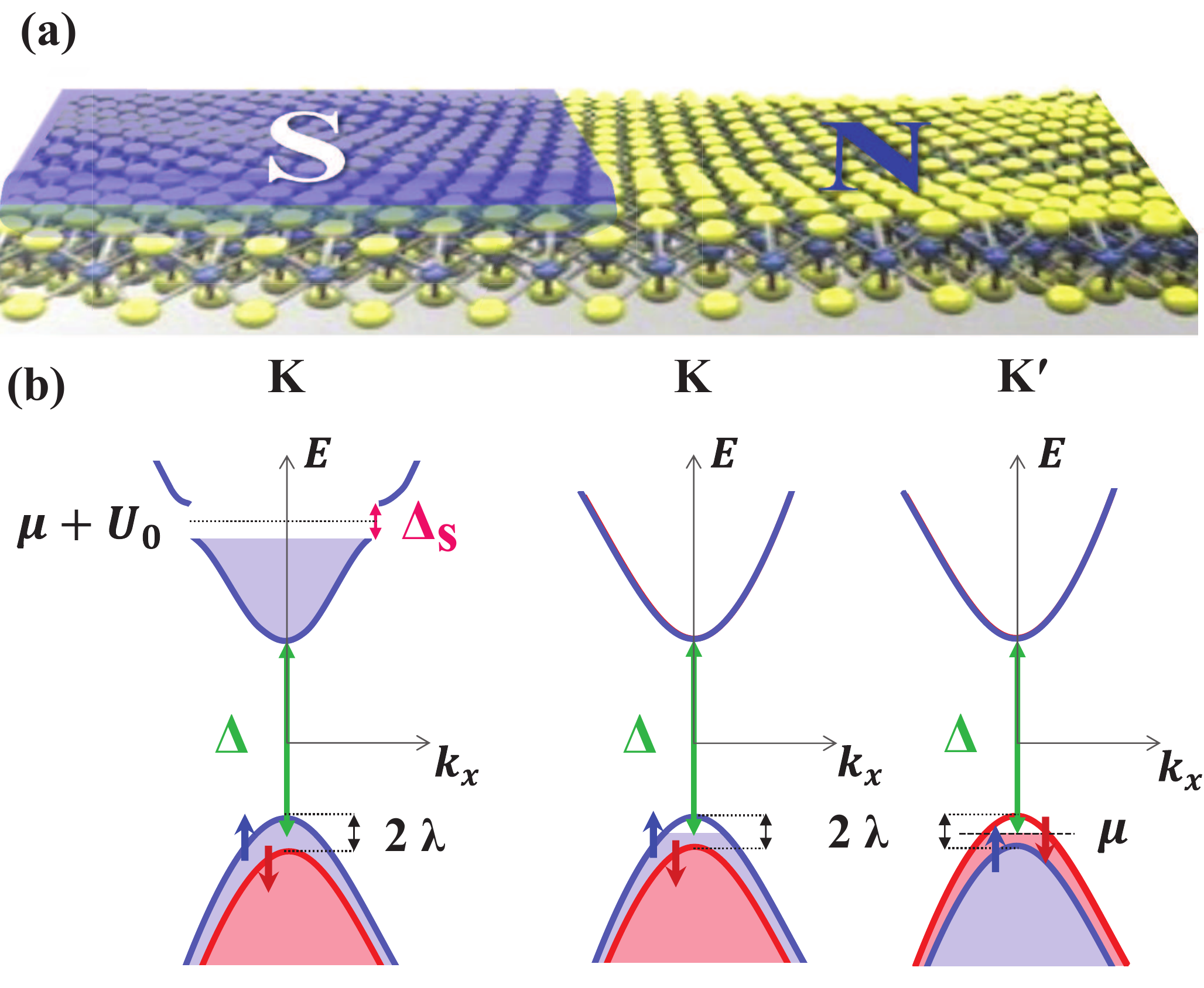}
\end{center}
\caption{\label{Fig:1} (Color online) (a) Schematic illustration of the molybdenum disulfide S/N junction. (b) The dispersion relation in momentum space ($E=E(k_x)$) of $n$-doped S (left panel) and $p$-doped N regions at $K$ and $K'$ valleys (middle and right panels). The conduction and the valence bands are separated by a large band gap $\Delta$. The strong spin-orbit coupling $\lambda$ splits the valence band to spin-up and spin-down subbands. The chemical potential $\mu$ is measured from the center of the gap $\Delta$ (zero-energy point).}
\end{figure}
Since the discovery of graphene~\cite{Novoselov04,Novoselov05,Zhang05}, there has been a growing interest in atomically thin two-dimensional (2D) crystals for application in nanoelectronic and optoelectronic devices. Layered transition metal dichalcogenides represent another class of materials that can be shaped into 2D monolayers.~\cite{Mattheis73} Recently, it has been demonstrated that monolayer molybdenum disulfide (MoS$_2$), prototypical group VI dichalcogenide, shows a transition from an indirect band gap of 1.3 eV in a bulk structure to a direct band gap of 1.9 eV in the monolayer structure.~\cite{Mak10,Splendiani10,Korn11} This intrinsic semiconducting nature of MoS$_2$ is a major advantage over graphene (which has no intrinsic band gap) as a 2D channel material in field effect transistors. More recently, a MoS$_2$ transistor with room-temperature mobility about 200 cm$^2$/(V.s) has been appeared.~\cite{Radisavljevic11} As in graphene, the electronic structure of monolayer MoS$_2$, exhibits a valley degree of freedom indicating that the conduction and valence band edges consist of two degenerate valleys $(K,K')$ located at the corners of the hexagonal Brillouin zone. Monolayer MoS$_2$ has important distinctions from graphene, too. Inversion symmetry is explicitly broken in monolayer MoS$_2$ which results in a strong coupling of the spin and valley degrees of freedom. MoS$_2$ has a strong spin-orbit coupling (originated from the heavy metal atoms) which splits the valence band to spin-up and spin-down subbands and leads to a spin polarization of the valence band.~\cite{Xiao12,Zeng12,Cao12} Recently, many measurements have been performed to characterize the optical and transport properties of the monolayer molybdenum disulfide.~\cite{Splendiani10,Radisavljevic11,Wang12,Sallen12} One important aspect of this material is that its potentialities for device applications are intimately related with fundamental concepts of quantum mechanics.
\par
Andreev reflection (AR) is a type of particle scattering which occurs at the interface between a normal metallic (N) and a superconducting (S) region.~\cite{Andreev64} In this process, an electron with an energy $\varepsilon$ (relative to the chemical potential $\mu$) and spin polarization $s$, upon hitting the N/S interface, is retro reflected as a hole with the same energy but opposite spin direction $-s$. This peculiar scattering process provides a conversion of the dissipative electrical current in N region into a dissipationless supercurrent and results in a finite conductance of N/S junction at the bias voltages below the superconducting energy gap $\varepsilon<\Delta_S$.~\cite{Blonder82} Novel interesting phenomena arise when N/S proximity structures are realized in atomically thin 2D crystals. Peculiarity of AR in graphene N/S junctions has been explained by Beenakker, who predicted the possibility for a specular AR in undoped normal graphene, and its associated anomaly in Andreev current-voltage characteristics of a graphene N/S contact.~\cite{beenakker06,beenakker08} Recently, another peculiarity of AR has been demonstrated in graphene-based superconducting hybrid structures, which is resulted from the sublattice pseudospin degree of freedom of electrons in graphene with a (non-superconducting) gap in its electronic spectrum.~\cite{majidi12,majidi13}
\par
In this paper, we theoretically study the superconducting proximity effect and focus on the signature of the AR process in a monolayer molybdenum disulfide S/N junction with $n$- ($p$-)doped S and $p$-doped N regions as sketched in Fig.~\ref{Fig:1}(a). Although $n$-type transistor operation for single-layer and few-layer MoS$_2$ with gold source and drain contacts was recently demonstrated~\cite{Radisavljevic11}, a multi-layer MoS$_2$ channel can be hole-doped by palladium contacts, yielding MoS$_2$ $p$-type transistors.~\cite{Fontana13} Recent studies~\cite{Gupta91, Takagi12, Ye12, Roldan13} have pointed out that MoS$_2$ undergoes a superconducting transition at high carrier concentration and in the presence of the high-$\kappa$ dielectrics, with a doping dependent temperature. We find that the electron-hole conversion with unit probability happens at normal incidence to the S/N structure with $p$-doped S region for $|\mu_N|>|\mu_S|$ in spite of the mismatch in the Fermi wave lengths at the two sides of the interface, while there is no perfect AR in the corresponding structure with $n$-doped S region. Furthermore, due to the spin-splitting of the valence band in the presence of the strong spin-orbit interaction, the AR process can be spin-valley polarized depending on the magnitude of the chemical potential $\mu_N$ and the excitation energy $\varepsilon$. The strong spin-orbit interaction enhances the probability of AR and the resulting Andreev conductance of the MoS$_2$-based S/N structure, relative to its value in the corresponding structure with gapped graphene. We further investigate the effect of the topological terms in the Hamiltonian of MoS$_2$~\cite{Rostami13} and show that the presence of the mass asymmetry term does not change the results significantly. More importantly, we show that the presence of $\beta$ term enhances the Andreev conductance of S/N structure with $p$-doped S region, while it reduces the Andreev conductance in the corresponding structure with $n$-doped S region. This is originated from the dependence of the probability of AR and the resulting Andreev conductance on the sign of $\beta$ and the chemical potential $\mu_S$. Accordingly, the presence of $\beta$ term amplifies the proximity-induced superconductivity inside the N region of the S/N structure with $p$-doped S region, in contrast to the S/N structure with $n$-doped S region.
\par
This paper is organized as follows. In Sec.~\ref{sec:level1}, we establish the theoretical framework which will be used to investigate AR in MoS$_2$-based S/N junction. We present our analytic and numeric results for the probabilities of the normal and Andreev reflections and the resulting Andreev conductance in Sec.~\ref{sec:level2}. Finally, a brief summary of results is given in Sec.~\ref{sec:level3}.
%%%%%%%%%%%%%%%%%%%%%%%%%%%%%%%%%%%%%%%%%%%%%%%%%%%%%%%%%%%%%%%%%%%%%%%%%%%
\section{\label{sec:level1}Model and Theory}
We consider a wide monolayer molybdenum disulfide (MoS$_2$) S/N hybrid junction with $n$- ($p$-)doped superconducting (S) region for $x<0$ and $p$-doped normal (N) region for $x>0$ as sketched in Fig.~\ref{Fig:1}(a). The superconducting part can be produced by depositing an S electrode on top of the MoS$_2$ sheet. In S region, the superconducting correlations are characterized by the superconducting pair potential (order parameter) $\Delta_S$ which is taken to be real and constant for s-wave pairing.
To study AR at S/N interface, we first derive Bogoliubov-de Gennes (BdG) equation for the system and then we construct the basis of scattering states in N and S regions. By introducing an effective mean-field Hamiltonian:
\begin{eqnarray}
H_{eff}&=&\int d\bm{r}\sum_s\ [\Psi^{\dagger}(\bm{r},s)H_0 \Psi(\bm{r},s)+U(\bm{r}) \Psi^{\dagger}(\bm{r},s) \Psi(\bm{r},s)]\nonumber\\
&+&[\Delta_S\Psi^{\dagger}(\bm{r},\uparrow)\Psi^{\dagger}(\bm{r},\downarrow)+\Delta^{\ast}_S\Psi(\bm{r},\downarrow)\Psi(\bm{r},\uparrow)]
\end{eqnarray}
and using the Bogoliubov transformations~\cite{De Gennes}, we obtain BdG equation as
\begin{eqnarray}
&&\hspace{1.5cm}H_{BdG}\left(
\begin{array}{c}
u\\
v
\end{array}
\right)
=\varepsilon\left(
\begin{array}{c}
u\\
v
\end{array}
\right),\nonumber\\\nonumber\\
\label{BdG}
&&\left(
\begin{array}{cc}
\mathcal{H}-\mu & \Delta_S \\
\Delta_{S}^{\ast}& \mu-\mathcal{T}\mathcal{H}\mathcal{T}^{-1}
\\
\end{array}
\right)
\left(
\begin{array}{c}
u\\
v
\end{array}
\right)
=\varepsilon\left(
\begin{array}{c}
u\\
v
\end{array}
\right),
\end{eqnarray}
which describes the superconducting correlations between electrons and holes with the wave functions $u$ and $v$. Here, $\mathcal{H}=H_0+U(\bm{r})$ is the effective single-particle Hamiltonian in the presence of an electrostatic potential $U(\bm{r})$, $\mathcal{T}$ is the time-reversal operator and $\varepsilon$ is the excitation energy.
\par
The effective single-particle Hamiltonian in monolayer MoS$_2$, where we ignore the intravalley interaction, is
\begin{equation}
\mathcal{H}=\left(
    \begin{array}{cc}
      \mathcal{H}_{\tau} & 0 \\
      0 & \mathcal{H}_{\bar{\tau}} \\
    \end{array}
  \right)
\end{equation}
where
\begin{equation}
\label{H}
\mathcal{H}_{\tau}=v_{\rm F}(\bm{\sigma}_{\tau}.\bm{p})+\frac{\Delta}{2}\sigma_{z}+\lambda s \tau\ (\frac{1-\sigma_z}{2})+\frac{\bm{p}^2}{4m_0}(\alpha+\beta\sigma_z)+U(\bm{r})
\end{equation}
\\
is the modified-Dirac Hamiltonian~\cite{Rostami13} for spin $s=\pm 1$ and valley $\tau=\pm 1$ ($\bar{\tau}=-\tau$) in which the energy gap $\Delta= 1.9$ eV, spin-orbit coupling constant $\lambda= 80$ meV, $v_{\rm F}=a_0t_0/\hbar=0.53\times10^6$ m/s is the Fermi velocity, $m_0$ is the bare electron mass, $\alpha = 0.43$ and $\beta=2.21$. The electrostatic potential $U(\bm{r})$ is taken to be $-U_0$ in S region and $U(\bm{r})=0$ in N region, and $\bm{\sigma}_{\tau}=(\tau\sigma_x,\sigma_y,\sigma_z)$ is the vector of the Pauli matrices acting on the two conduction and valence bands. Introducing the time-reversal operator as
\begin{equation}
\mathcal{T}=i\tau_x s_y \mathcal{K},
\end{equation}
with $\mathcal{K}$ the operator of complex conjugation, we obtain that the Hamiltonian of MoS$_2$ is time-reversal invariant $\mathcal{T}\mathcal{H}(\bm{p})\mathcal{T}^{-1}=\mathcal{H}(-\bm{p})$. Since the doped monolayer molybdenum disulfide is a Fermi liquid, the superconducting states can be established for the system and thus the BdG Hamiltonian (Eq. \ref{BdG}) has particle-hole symmetry $(\mathcal{T}C) H_{BdG}(\bm{p})+H_{BdG}(\bm{p})(\mathcal{T}C)=0$ with $C=i\gamma_y$, in which $C H_{BdG}(\bm{p})C^{-1}=-H_{BdG}(-\bm{p})$ (the Pauli matrix $\gamma_y$ acts on the electron-hole space). Substituting the time-reversal operator $\mathcal{T}$ into Eq. (\ref{BdG}), results in two decoupled sets of four-dimensional Dirac-Bogoliubov-de Gennes (DBdG) equations~\cite{beenakker06}, which each of the form is given by
\begin{equation}
\label{DBdG}
\hspace{-0.5cm}\left(
\begin{array}{cc}
\mathcal{H}_{\tau}-\mu & \Delta_S \\
\Delta_{S}^{\ast}& \mu-\mathcal{H}_{\tau}
\\
\end{array}
\right)
\left(
\begin{array}{c}
u_{\tau}\\
v_{\bar{\tau}}
\end{array}
\right)
=\varepsilon\left(
\begin{array}{c}
u_{\tau}\\
v_{\bar{\tau}}
\end{array}
\right).
\end{equation}
The electron and hole wave functions, $u_{\tau}$ and $v_{\bar{\tau}}$, are two-component spinors of the form $(\psi_c,\psi_v)$, where $c$ and $v$ denote the conduction and valence bands, respectively. Therefore, the electron excitations in one valley are coupled by the superconducting pair potential $\Delta_{S}$ to hole excitations in the other valley.
\par
The solutions of DBdG equation inside the S region are rather mixed electron-hole excitations (called Dirac-Bogoliubov quasiparticles) that either decay exponentially as $x\rightarrow -\infty$ (for subgap solutions when $\varepsilon<\Delta_S$) or propagate along the $-x$ direction (for supragap solutions when $\varepsilon>\Delta_S$). These solutions for $n$-doped S region take the form
\begin{eqnarray}
\psi^{S+}=e^{ ik'_{+}\tau x} e^{iqy}
\left(
\begin{array}{c}
  b_{+}^{-1} \\
  -a_{+}\ c_{+}^{-1} \\
  1\\
  -a_{+}
\end{array}
\right),
\\\nonumber\\
\psi^{S-}=e^{ ik'_{-}\tau x} e^{iqy}
\left(
\begin{array}{c}
  b_{-}^{-1} \\
  -a_{-}\ c_{-}^{-1} \\
  1\\
  -a_{-}
\end{array}
\right),
\end{eqnarray}
where
\begin{eqnarray}
&&\hspace{-5mm}a_{\pm}=\frac{m_{\pm}+{\Delta_S^2}}{\hbar v_{\rm F} (\tau k'_{\pm}-iq)\ m'_{\pm}},\nonumber\\
&&\hspace{-5mm}b_{\pm}=\frac{-m'_{\pm}\ (\frac{\Delta}{2}+\frac{\hbar^2 k_{S\pm}^2 (\alpha+\beta)}{4m_0}-\mu_S-\varepsilon)+ m_{\pm}+{\Delta_S^2}}{m'_{\pm}\ \Delta_S},\nonumber\\
&&\hspace{-5mm}c_{\pm}=\nonumber\\
&&\hspace{-5mm}\frac{\hbar^2v_{\rm F}^2{k}_{S\pm}^2 m'_{\pm}-(m_{\pm}+{\Delta_S^2})(-\frac{\Delta}{2}+\lambda s \tau+\frac{\hbar^2 k_{S\pm}^2 (\alpha-\beta)}{4m_0}-\mu_S-\varepsilon)}{\Delta_S\ (m_{\pm}+{\Delta_S^2})},\nonumber\\\nonumber
\end{eqnarray}
$k'_{\pm}=\pm k_0-i\kappa\tau$, $k_{S\pm}=\sqrt{{k'_{\pm}}^2+q^2}$, $\mu_S=\mu+U_0$ ($\mu_S$ is measured from the center of the gap $\Delta$), $m_{\pm}=({\Delta}/{2}+\hbar^2 k_{S\pm}^2 (\alpha+\beta)/4m_0-\mu_S)^2-\varepsilon^2+\hbar^2v_{\rm F}^2{k}_{S\pm}^2$ and $m'_{\pm}=\hbar^2 k_{S\pm}^2 \alpha/2m_0-2\mu_S+\lambda s \tau$. The momentum $k_{S\pm}$ of the qausiparticles in S region are the solutions of the energy-momentum relation, which can be obtained by solving the following equation
\begin{widetext}
\begin{eqnarray}
\label{ks}
&&\hspace{-5mm}\varepsilon^4 - d\ \varepsilon^2 + f = 0,\\
&&\hspace{-5mm}d=\left(\frac{\hbar^2 k_S^2 }{4 m_0}(\beta -\alpha )-\lambda s \tau +\frac{\Delta }{2}+\mu_S\right)^2+\left(\frac{\hbar^2 k_S^2 }{4m_0}(\alpha +\beta )+ \frac{\Delta }{2} - \mu_S\right)^2+2\ (\hbar^2 v_{\rm F}^2 k_S^2+ \Delta_S^2),\nonumber\\
&&\hspace{-5mm}f=\left([\frac{\hbar ^2 {k_S}^2 }{4 m_0}(\alpha +\beta )+\frac{\Delta }{2}-\mu_S]^2+\hbar ^2 v_{\rm F}^2 {k_S}^2+\Delta_S^2\right) \left([\ \frac{\hbar ^2 {k_S}^2}{4 m_0} (\beta-\alpha)- \lambda s \tau +\frac{\Delta }{2}+\mu_S ]^2+\hbar ^2 v_{\rm F}^2 {k_S}^2+\Delta_S^2\right)\nonumber\\
&&\hspace{0.2cm}-\hbar ^2 v_{\rm F}^2 {k_S}^2 \left(\frac{\hbar ^2 {k_S}^2}{2m_0}\ \alpha + \lambda s  \tau -2 \mu_S\right)^2.\nonumber
\end{eqnarray}
\end{widetext}
\par
Inside N region, the solutions of DBdG equation are two states of the form
\begin{equation}
\label{psie}
\psi^{e\pm}=\frac{1}{\sqrt{u_e}}\ e^{\mp ik_e\tau x} e^{iqy}
\left(
\begin{array}{c}
e^{\pm i\tau\theta_e/2}\\
\mp A_e\ \tau\ e^{\mp i\tau\theta_e/2}\\
0\\
0
\end{array}
\right),
\end{equation}
for the valence band electrons and
\begin{equation}
\label{psih}
\psi^{h\pm}=\frac{1}{\sqrt{u_h}}\ e^{\pm ik_h\tau x} e^{iqy}
\left(
\begin{array}{c}
0\\
0\\
e^{\mp i\tau\theta_h/2}\\
\pm A_h\ \tau\ e^{\pm i\tau\theta_h/2}
\end{array}
\right),
\end{equation}
for the valence band holes of $p$-doped MoS$_2$ with $\mu_N=\mu\ $ ($\mu<0$ is measured from the center of the gap $\Delta$), at a given energy $\varepsilon$ and a transverse momentum $q$ with energy-momentum relations that can be obtained by solving the following equation,
\begin{widetext}
\begin{equation}\label{general}
\left(\frac{\hbar^2|\bm{k}_{e(h)}|^2}{4 m_0} (\alpha +\beta )+\frac{\Delta }{2}-\mu_N\mp\epsilon \right)\ \left(\frac{\hbar^2 |\bm{k}_{e(h)}|^2 }{4 m_0}(\alpha -\beta )+\lambda s  \tau -\frac{\Delta }{2}-\mu_N\mp\epsilon \right)-\hbar^2 v_{\rm F}^2 |\bm{k}_{e(h)}|^2=0.
\end{equation}
\end{widetext}
In Eqs.~(\ref{psie}) and (\ref{psih}), $u_{e(h)}=\hbar |\bm{k}_{e(h)}|\cos({\tau\theta_{e(h)}})\ [\alpha+\beta+A_{e(h)}^2\ (\alpha-\beta)]/4m_0 v_{\rm F}+A_{e(h)}\cos({\tau\theta_{e(h)}})$, $A_{e(h)}=\hbar v_{\rm F} |\bm{k}_{e(h)}|/[\mu_N\pm\varepsilon+\Delta/2-\lambda s \tau-\hbar^2|\bm{k}_{e(h)}|^2(\alpha-\beta)/4 m_0]$ and $\theta_{e(h)}=\arcsin({q/|\bm{k}_{e(h)}|})$ indicates the angle of propagation of the electron (hole). Also, the two propagation directions along the $x$-axis are denoted by $\pm$ in $\psi^{e(h)\pm}$.
\par
An incoming electron from the valence band of $p$-type N region with a subgap energy $\varepsilon\leq\Delta_S$ may be either normally reflected as an electron or Andreev reflected as a hole in the same band (retro reflection). Due to the spin-splitting of the valence band, the incident electron and the reflected hole can be from one or two of the spin subbands, depending on the magnitude of the chemical potential $\mu_N$ and the excitation energy $\varepsilon$. As can be seen from Fig.~\ref{Fig:1}(b), as long as $-\Delta/2-\lambda+\varepsilon<\mu_N\leq-\Delta/2+\lambda+\varepsilon$ only the upper spin subbands with $s=\tau=1$ and $s=\tau=-1$ contribute to the transport of charge and result in a spin-valley polarized AR process with $s\tau=1$. While for the case of $\mu_N\leq-\Delta/2-\lambda+\varepsilon$, the Fermi level crosses the two spin subbands with $s\tau=\pm1$ and therefore the AR process is not spin-valley polarized.
\par
From the conservation of the $y$-component wave vector $q$ under the scattering process, we obtain the following relation between the incident electron and reflected hole angels,
\begin{equation}
\label{qconservation}
{|\bm{k}_e|}\sin{\theta_e}={|\bm{k}_h|}\sin{\theta_h}.
\end{equation}
Denoting the amplitudes of normal and Andreev reflections, $r^{s,\tau}$ and $r_{A}^{s,\tau}$, respectively, the wave functions inside N and S regions are written as
\begin{eqnarray}
\label{N wave function}
&&\psi_{N}=\psi^{e-}+r^{s,\tau}\ \psi^{e+}+r_{A}^{s,\tau}\ \psi^{h+},\\
\label{s wave function}
&&\psi_{S}=t\ \psi^{S+}+t'\ \psi^{S-}.
\end{eqnarray}
Matching the wave functions of N and S regions at the interface $x=0$, we obtain
\begin{widetext}
\label{r}
\begin{eqnarray}
\label{rr}
\hspace{-1.5cm}r^{s,\tau}&=&\frac{a\ a'\ b\ b'(c'- c)+ A_e\ \tau\ c\ c'e^{i \theta_e \tau} (a'\ b'-a\ b) - A_h\ \tau e^{i \theta_h \tau} \ [b\ b'(a'\ c - a\ c')+ A_e\ \tau\ c\ c'(b - b') e^{i \theta_e \tau}]}{-a\ b\ [-a'\ b'\ (c - c') e^{i \theta_e \tau} + c'\ \tau(A_e\ c + A_h\ b' e^{i (\theta_e + \theta_h) \tau})] + c\ \tau [A_h\ a'\ b\ b'\ e^{i (\theta_e+\theta_h) \tau} + A_e\ c' (a'\ b' + A_h (b'-b)\tau e^{i \theta_h \tau})]},\nonumber\\\nonumber\\\\
\label{rrA}
\hspace{-1.5cm}r_{A}^{s,\tau}&=&\frac{-A_e\  (a - a')\ b\ b'\ c\ c'\ e^{-i(\theta_e - \theta_h ) \tau/2 } (1 + e^{2 i \theta_e \tau})\ \tau\sqrt{u_h/u_e}}{-a\ b\ [-a'\ b'\ (c - c') e^{i \theta_e \tau} + c'\ \tau(A_e\ c + A_h\ b' e^{i (\theta_e + \theta_h) \tau})] + c\ \tau [A_h\ a'\ b\ b'\ e^{i (\theta_e+\theta_h) \tau} + A_e\ c' (a'\ b' + A_h (b'-b)\tau e^{i \theta_h \tau})]}.\nonumber\\
\end{eqnarray}
\end{widetext}
Having obtained the above reflection amplitudes, we could analysis the Andreev conductance of a S/N interface with $n$- ($p$-)doped S and $p$-doped N regions and the results will be discussed in the next section.
\begin{figure}[]
\begin{center}
\includegraphics[width=1\linewidth]{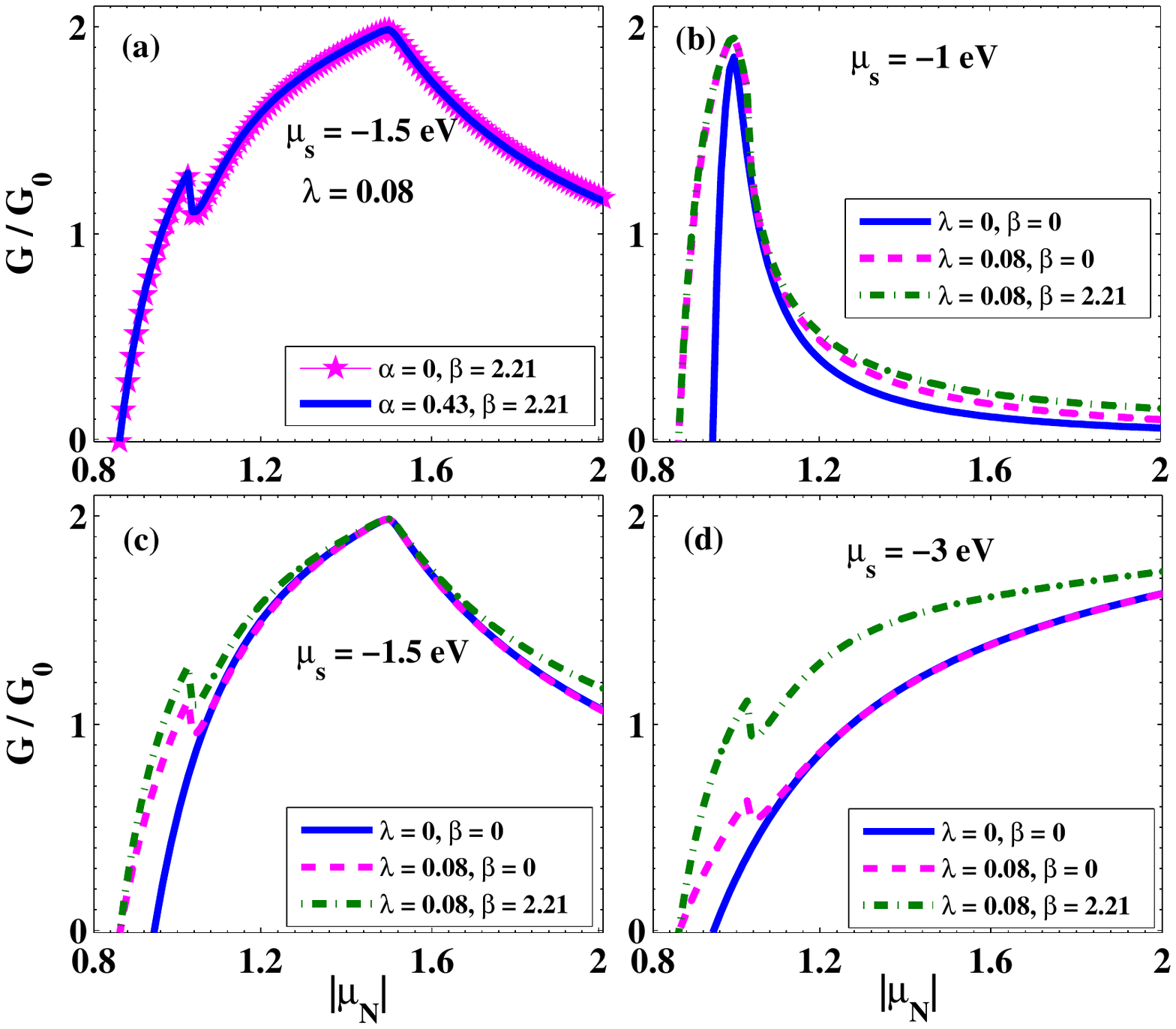}
\end{center}
\caption{\label{Fig:2}(Color online) Dependence of the Andreev conductance of S/N junction with $p$-doped S and N regions on the magnitude of the chemical potential $|\mu_N|$ (in units of eV) for MoS$_2$-based structure ($\lambda=0.08$ eV) with $\mu_S = -1.5$ eV, $\beta=2.21$, and $\alpha=0$ and $0.43$ (a) and for gapped graphene ($\lambda=0, \beta=0$), MoS$_2$ with $\alpha=0$ and $\beta=0$ and $2.21$ at three different chemical potentials $\mu_S = -1,-1.5$ and $-3$ eV (b-d), when $\Delta_S = 0.01$ eV and $\varepsilon/\Delta_S=eV/\Delta_S = 0$.}
\end{figure}
%%%%%%%%%%%%%%%%%%%%%%%%%%%%%%%%%%%%%%%%%%%%%%%%%%%%%%%%%%%%%%%%%%%%%%%%%%%%%%%%%%%%%%%%%%%%%%%%%%%%%%%%%%%%
\section{\label{sec:level2}NUMERICAL RESULTS AND DISCUSSIONS}
\begin{figure}[t]
\begin{center}
\includegraphics[width=1\linewidth]{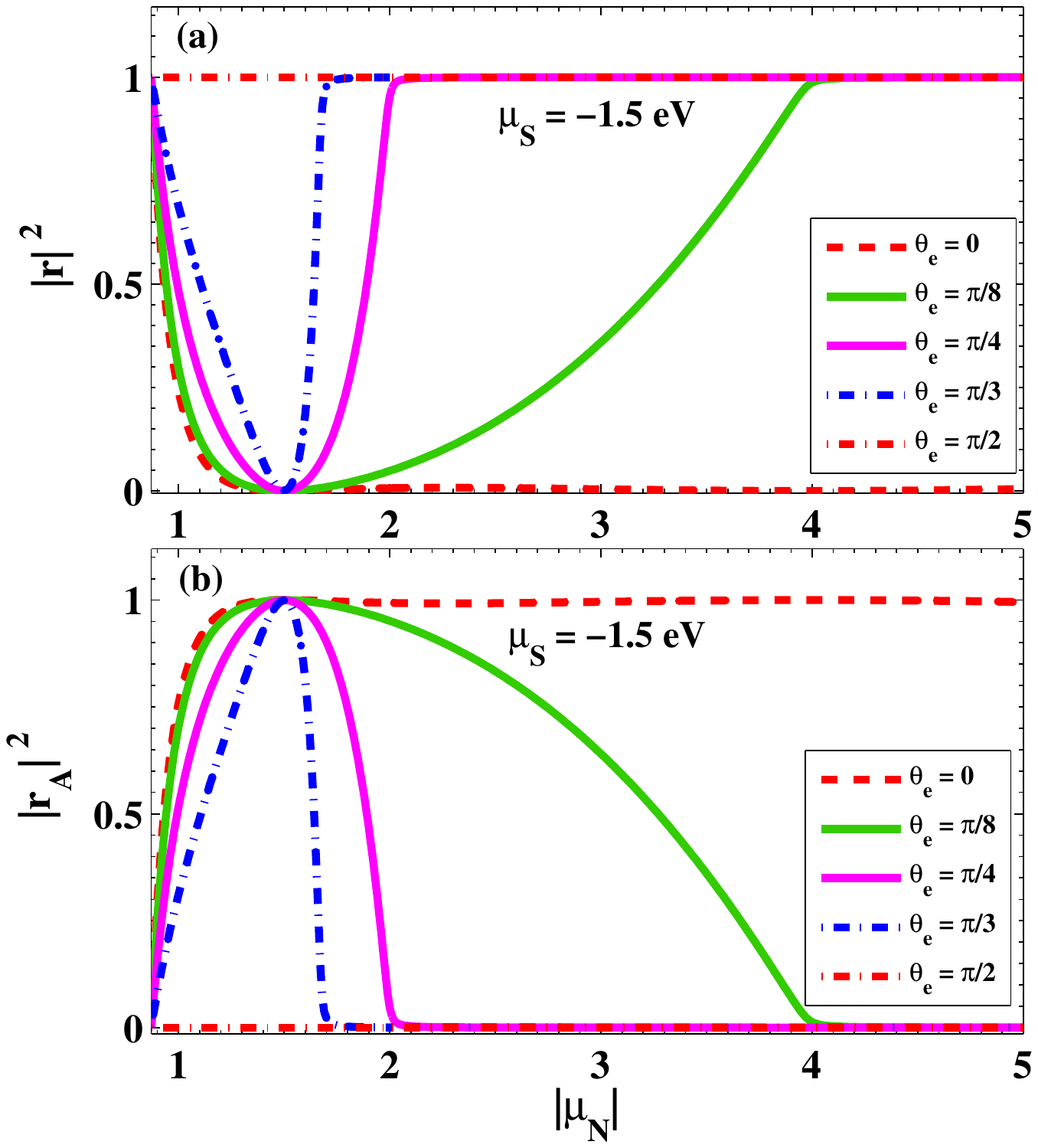}
\end{center}
\caption{\label{Fig:3} (Color online) Normal and Andreev reflection probabilities as a function of the chemical potential $|\mu_N|$, for different angles of incidence to the MoS$_2$-based S/N junction with $p$-doped S and N regions, when $\mu_S=-1.5$ eV, $s=\tau=1$, $\alpha=0$, $\beta=2.21$ and $eV/\Delta_S = 0$.}
\end{figure}
To evaluate the Andreev differential conductance of a MoS$_2$-based S/N structure at zero temperature, we use the Blonder-Tinkham-Klapwijk (BTK) formula\cite{Blonder82} which is given by
\begin{equation}
\label{G}
G=\sum_{s,\tau=\pm1}G_{0}^{s,\tau}\int_{0}^{\theta_{c}}(1-|r^{s,\tau}|^2+|r_{A}^{s,\tau}|^2)\cos{\theta_e}\ d\theta_e,
\end{equation}
where we introduce
\begin{equation}
G_{0}^{s,\tau}=\frac{e^2}{h}N_{s,\tau}(eV),
\end{equation}
as the spin-$s$ valley-$\tau$ normal state conductance and $\ N_{s,\tau}(\varepsilon)={W|\bm{k}_e|}/{\pi}$ as the number of transverse modes in a sheet of monolayer MoS$_2$ of width W. Here, $\theta_{c}=\arcsin({|\bm{k}_h|}/{|\bm{k}_e|})$ is the critical angle of incidence above which the Andreev reflected waves become evanescent and do not contribute to any transport of charge. Also, we have put $\varepsilon=eV$ at zero temperature. We note that in contrast to the valley degeneracy in graphene, the contribution of each valley to the charge conductance must be computed separately.
\par
We present our numerical results, obtained using the numerical $r^{s,\tau}$, $r_A^{s,\tau}$ and $G$ based on Eqs.~(\ref{general}), (\ref{rr}), (\ref{rrA}) and (\ref{G}), in the physical regime. We first discuss the AR process for the S/N structure with $p$-doped N and S regions and then the case with $n$-doped S region.
\subsection{\label{subsec:1}S/N structure with $p$-doped S and N regions}
The resulting Andreev conductance of S/N structure $G/G_0$ ($G_0=\sum_{s,\tau=\pm1}G_{0}^{s,\tau}$) with $p$-doped S and N regions is presented in Fig.~\ref{Fig:2} in terms of the magnitude of the chemical potential $|\mu_N|$ for MoS$_2$-based structure ($\lambda=0.08$ eV) with $\beta=2.21$, and $\alpha=0$ and $0.43$ (Fig. \ref{Fig:2}(a)) and also for gapped graphene ($\lambda=0, \beta=0$), MoS$_2$ with $\alpha=0$ and $\beta=0$ and $2.21$ at three different chemical potentials $\mu_S = -1, -1.5$ and $-3$ eV (Figs.~\ref{Fig:2}(b-d)), when $\Delta_S = 0.01$ eV and $\varepsilon/\Delta_S=eV/\Delta_S = 0$. As can be seen from Fig.~\ref{Fig:2}(a), the presence of the mass asymmetry term ($\alpha$) in the Hamiltonian, which is originated from the difference between electron and hole masses~\cite{Rostami13,Peelaers12}, has no effect on the magnitude of the Andreev conductance in MoS$_2$-based structure. Therefore in the following investigations, we put $\alpha=0$. Also, it is seen from Figs.~\ref{Fig:2}(a-d) that the absence of quasiparticle states inside the band gap of N region causes a gap in conductance for $|\mu_N|<|-\Delta/2+\lambda|$. For $|-\Delta/2+\lambda|\leq|\mu_N|\leq|\mu_S|$ (with negative $U_0$), the zero bias Andreev conductance increases with $|\mu_N|$ and reaches a maximum value at $|\mu_N|=|\mu_S|$ where the electrostatic potential $U_0$ induced by the superconducting electrode is zero, while for $|\mu_N|>|\mu_S|$ (with positive $U_0$) the conductance decreases by increasing $|\mu_N|$.
\par
In order to explain the behavior of the Andreev conductance, we plot the $|\mu_N|$ dependence of the normal and Andreev reflection probabilities in Fig. \ref{Fig:3} for different angles of incidence, when $\mu_S = -1.5$ eV, $s=\tau=1$, $\alpha=0$, $\beta=2.21$ and $eV/\Delta_S = 0$. For $|\mu_N|\leq|\mu_S|$, the magnitude of the negative electrostatic potential $U_0$ and therefore the probability of the normal reflection decreases with $|\mu_N|$ and goes to zero at $|\mu_N|=|\mu_S|$. Therefore, the probability of AR and the resulting Andreev conductance increase with $|\mu_N|$ and the electron-hole conversion with unit probability happens at $|\mu_N|=|\mu_S|$. While for $|\mu_N|>|\mu_S|$, the positive electrostatic potential increases with $|\mu_N|$ and leads to the increasing behavior of the normal reflection probability. So the probability of AR and the resulting Andreev conductance decrease with $|\mu_N|$.
\par
\begin{figure}[t]
\begin{center}
\includegraphics[width=1\linewidth]{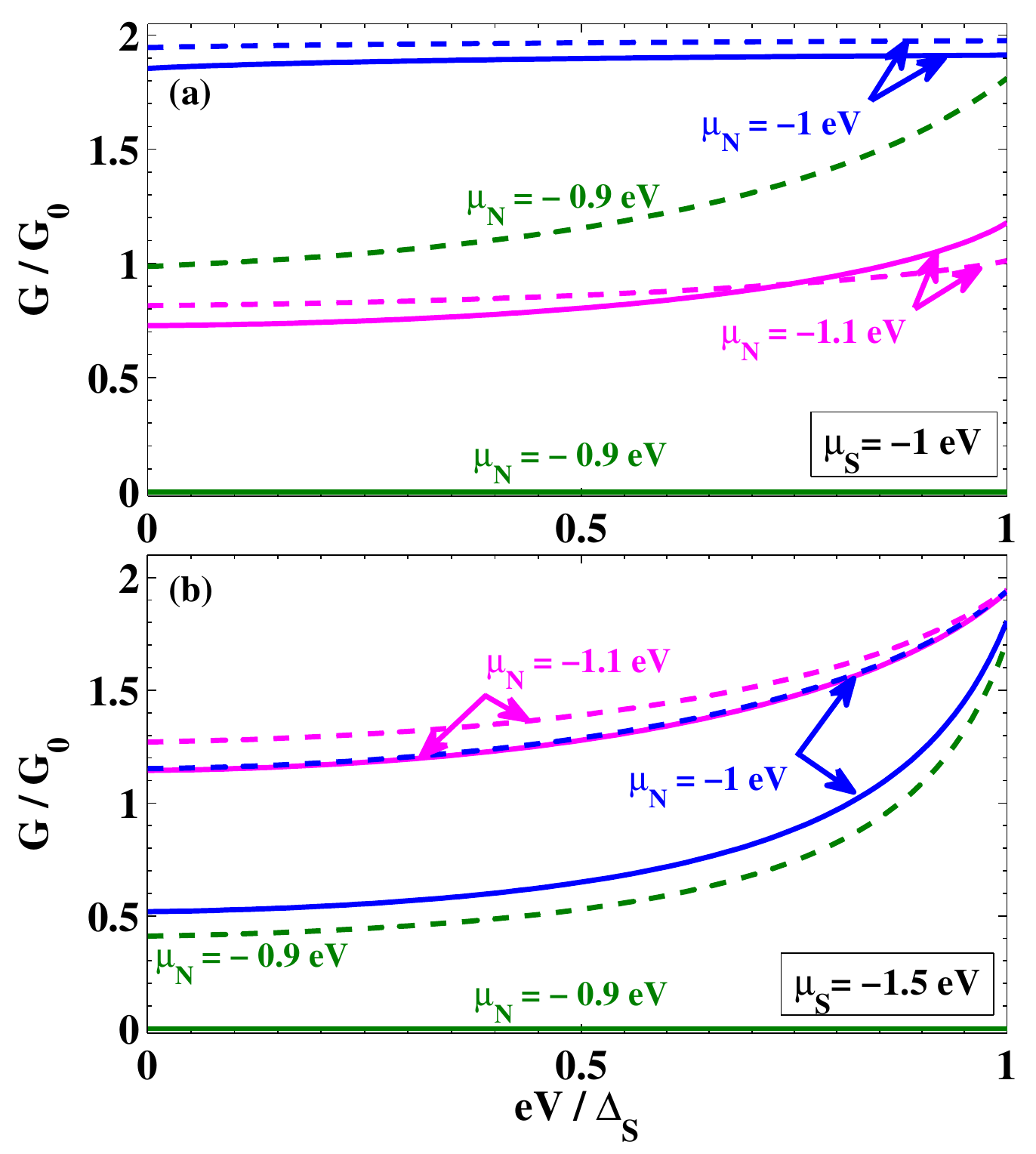}
\end{center}
\caption{\label{Fig:4} (Color online) Andreev conductance of S/N junction with $p$-doped S and N regions as a function of the bias voltage $eV/\Delta_S$ (in units of the superconducting gap $\Delta_S$) at two values of $\mu_S = -1$ and $-1.5$ eV, for gapped graphene (solid lines) and MoS$_2$ with $\alpha=0$ and $\beta=2.21$ (dashed lines), when $\mu_N = -0.9, -1$ and $-1.1$ eV.}
\end{figure}
Moreover, due to the spin-splitting effect of the spin-orbit interaction, a cusp-like behavior is appeared at the edge of the spin-down (-up) subband of $K$ ($K'$) valley with $|\mu_N|=|-\Delta/2-\lambda|$ such that the AR process for $|\mu_N|<|-\Delta/2-\lambda|$ is spin-valley polarized. Importantly, we see that the presence of the strong spin-orbit interaction and $\beta$ term in the Hamiltonian enhances the amplitude of AR and the resulting Andreev conductance in MoS$_2$-based S/N structure, as is compared with its value in gapped graphene S/N structure. Furthermore, the bias voltage dependence of the Andreev conductance of S/N structure for gapped graphene (solid lines) and MoS$_2$ with $\alpha=0$ and $\beta=2.21$ (dashed lines) are shown in Fig. \ref{Fig:4} at two values of $\mu_S = -1$ and $-1.5$ eV, when $\mu_N = -0.9, -1$ and $-1.1$ eV. It is seen that the Andreev conductance of both structures increases with the bias voltage $eV/\Delta_S$ (in units of the superconducting gap $\Delta_S$) for different values of the chemical potential $\mu_N$ with $|\mu_N|\geq|-\Delta/2+\lambda|$. Also we can see that the enhancement of the Andreev conductance of MoS$_2$-based structure in presence of the strong spin-orbit interaction and $\beta$ term in the Hamiltonian can be occurred for subgap bias voltages, depending on the magnitude of chemical potentials $\mu_N$ and $\mu_S$. Also we can see from Figs.~\ref{Fig:2}(b-d) that by enhancing the chemical potential of the S region ($\mu_S$), the magnitude of the Andreev conductance for MoS$_2$-based structure with $\beta=2.21$ increases more than that of the corresponding structure with zero $\beta$.
\par
\begin{figure}[t]
\begin{center}
\includegraphics[width=1\linewidth]{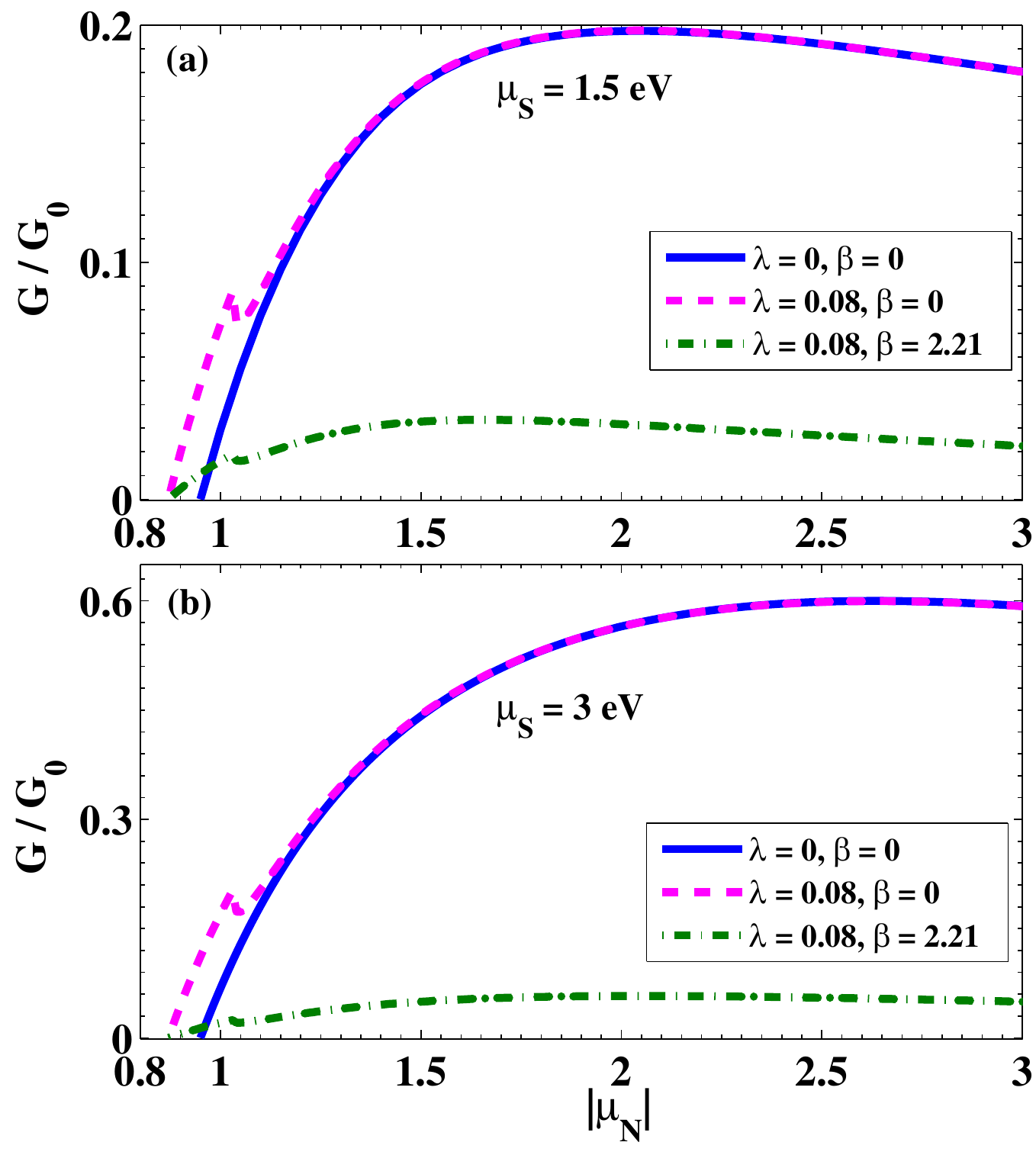}
\end{center}
\caption{\label{Fig:5} (Color online) Andreev conductance of S/N junction with $n$-doped S region as a function of $|\mu_N|$ at two chemical potentials $\mu_S = 1.5$ and $3$ eV, for gapped graphene ($\lambda=0, \beta=0$), MoS$_2$ ($\lambda=0.08$ eV) with $\alpha=0$ and $\beta=0$ and $2.21$, when $\Delta_S = 0.01$ eV and $eV/\Delta_S = 0$.}
\end{figure}
\begin{figure}[]
\begin{center}
\includegraphics[width=0.99\linewidth]{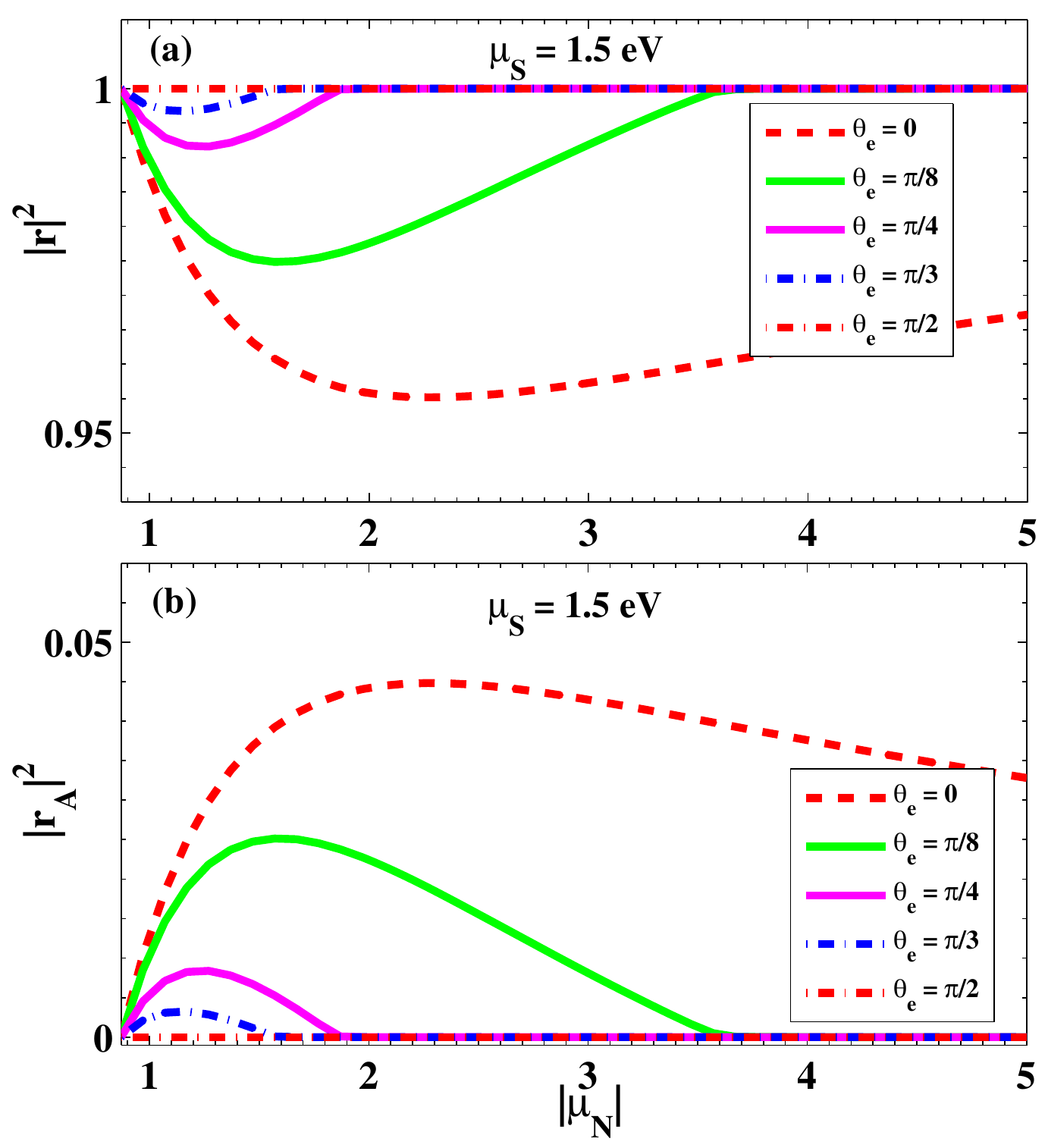}
\end{center}
\caption{\label{Fig:6} (Color online) Normal and Andreev reflection probabilities as a function of $|\mu_N|$, for different angles of incidence to the MoS$_2$-based S/N junction with $n$-doped S region, when $\mu_S=1.5$ eV, $s=\tau=1$, $\alpha=0$, $\beta=2.21$ and $eV/\Delta_S = 0$.}
\end{figure}
\subsection{\label{subsec:2}S/N structure with $n$-doped S and $p$-doped N regions}
The behavior of the Andreev conductance of S/N structure with $n$-doped S and $p$-doped N  regions is shown in Fig.~\ref{Fig:5} in terms of the chemical potential $|\mu_N|$ at two chemical potentials $\mu_S = 1.5$ and $3$ eV, when $\alpha=0$ and $eV/\Delta_S = 0$. It is seen that the Andreev conductance has an increasing behavior with $|\mu_N|$ and reaches a maximum at higher chemical potentials. This behavior can be explained in terms of the behavior of the normal and Andreev reflection probabilities, which are shown in Fig.~\ref{Fig:6}. It is seen that the normal reflection probability decreases with $|\mu_N|$ (positive $U_0$) and reaches a minimum at $\mu_0$, which can be obtained from Eq. (\ref{rr}). Therefore the probability of AR and the resulting Andreev conductance increase with $|\mu_N|$ and reach a maximum at $\mu_0$. While for $|\mu_N|>|\mu_0|$, the probability of AR and therefore the Andreev conductance decrease with $|\mu_N|$ where the normal reflection probability has an increasing behavior with $|\mu_N|$. This behavior of the Andreev conductance is similar to that of a interface between a $p$-type semiconductor and a conventional superconductor\cite{Futterer11}.
\par
Also it is seen from Fig.~\ref{Fig:5} that the presence of strong spin-orbit interaction in molybdenum disulfide enhances the Andreev conductance, while the presence of $\beta$ term in the Hamiltonian reduces the Andreev conductance of MoS$_2$-based S/N structure from its value for the corresponding gapped graphene structure. Moreover, the magnitude of the Andreev conductance for all three structures can be enhanced by $\mu_S$. Therefore, the presence of $\beta$ term in the Hamiltonian of monolayer molybdenum disulfide attenuates the proximity-induced superconductivity inside the N region of the S/N junction with $n$-doped S region, in contrast to the corresponding structure with $p$-doped S region.
\par
We obtain numerically that the amplitude of AR can be written as $|r_A(\beta)|=|r_A(0)|-\beta \operatorname{sgn}(\mu_S-E_b)\ |r'_A(0)|$ in the limit of small $\beta$, where $E_b$ indicates the conduction, $E_{CBM}$ (valence,  $E_{VBM}$) band edge. So the presence of $\beta$ term reduces the amplitude of AR for S/N structure with $n$-doped S region where $\mu_S>E_{CBM}$ and enhances it for the corresponding structure with $p$-doped S region where $\mu_S<E_{VBM}$. This result persists for finite $\beta$ and tells us that the probability of AR and the resulting Andreev conductance depend on the sign of the chemical potential $\mu_S$ in the superconducting region. Also, the magnitude of the Andreev conductance depends on the sign of $\beta$ and can be enhanced (reduced) for S/N structure with $n$- ($p$-)doped S region, when $\beta$ becomes negative.
\par
Furthermore, we have presented the behavior of the normal and Andreev reflection probabilities in terms of $|\mu_N|$ for different angles of incidence to the two S/N junctions in Figs. \ref{Fig:3} and \ref{Fig:6}. We see that the probabilities of the normal and Andreev reflections strongly depend on the angle of incidence $\theta_e$ such that the normal reflection probability increases with $\theta_e$ and totaly dominates AR, when the incident electron is parallel to the S/N interface. Moreover, depending on the angle of incidence, there is a critical chemical potential for N region (it can be obtained from Eq. \ref{qconservation}) above which the AR process suppresses and the normal reflection happens with unit probability. Also it is seen that for the S/N structure with $p$-doped S region, perfect electron-hole conversion with $|r_A|^2=1$ happens at normal incidence when $|\mu_N|\geq|\mu_S|$ and away from the normal incidence when $|\mu_N|=|\mu_S|$ (see Fig. \ref{Fig:3}(b)), while there is no perfect AR in the corresponding structure with $n$-doped S region (see Fig. \ref{Fig:6}(b)). The existence of the perfect AR for $p$-doped S/N structure with $|\mu_N|>|\mu_S|$, is in spite of the mismatch in Fermi wave lengths at the two sides of the S/N interface.
%%%%%%%%%%%%%%%%%%%%%%%%%%%%%%%%%%%%%%%%%%%%%%%%%%%%%%%%%%%%%%%%%%%%%%%%%%%%%%%%%%%%%%%%%%%%%%%%%%
\section{\label{sec:level3}Conclusion}
In conclusion, we have investigated the superconducting proximity effect and specially the Andreev reflection in a molybdenum disulfide superconducting-normal (S/N) hybrid junction with $n$- ($p$-)doped S and $p$-doped N regions. We have realized that the electron-hole conversion with unit efficiency happens at normal incidence to the S/N interface with $p$-doped S region when $|\mu_N|\geq|\mu_S|$. The presence of the strong spin-orbit coupling in molybdenum disulfide enhances the Andreev conductance of the MoS$_2$-based S/N structure relative to its value in the corresponding structure with gapped graphene. We have further analyzed the effect of the topological terms in the Hamiltonian of MoS$_2$, which have been reported in Ref.~\onlinecite{Rostami13}, and demonstrated that the presence of $\beta$ term results in an enhancement of the Andreev conductance of S/N structure with $p$-doped S region, while it reduces the Andreev conductance in the corresponding structure with $n$-doped S region. This effect is due to the dependence of the Andreev reflection probability on the sign of $\beta$ and the chemical potential in the superconducting region. Moreover, we have found that the presence of the mass asymmetry term in the Hamiltonian, does not change the results, significantly.
\par
The role of the finite-size effect for a nanoribbon molybdenum disulfide has not been addresses in the present work. We remark that, in the very large chemical potential regime, where the system is highly doped a model going beyond the low-energy modified-Dirac Hamiltonian is necessary to account the full dispersion relation.

\end{document}